# Ammonia-induced Calcium Phosphate Nanostructure: A Potential Assay for Studying Osteoporosis and Bone Metastasis

**Sijia Chen[a,c], Qiong Wang[a,b,c], Felipe Eltit[a,b,c], Yubin Guo[d], Michael Cox[d], Rizhi Wang[a,b,c,*]**

a. Department of Materials Engineering, University of British Columbia, Vancouver, BC, Canada.
b. School of Biomedical Engineering, University of British Columbia, Vancouver, BC, Canada.
c. Centre for Hip Health and Mobility, University of British Columbia, Vancouver, BC, Canada.
d. Vancouver Prostate Centre, University of British Columbia, Vancouver, BC, Canada.

**Abstract**: Osteoclastic resorption of bone plays a central role in both osteoporosis and bone metastasis. A reliable *in vitro* assay that simulates osteoclastic resorption *in vivo* would significantly speed up the process of developing effective therapeutic solutions for those diseases. Here we reported the development of a novel and robust nano-structured calcium phosphate coating with unique functions on the track-etched porous membrane by using an ammonia-induced mineralization (AiM) technique. The calcium phosphate coating uniformly covers one side of the PET membrane enabling testing for osteoclastic resorption. The track-etched pores in the PET membrane allow calcium phosphate mineral pins to grow inside, which, on one hand, enhances coating integration with membrane substrate, and on the other hand provides diffusion channels for delivering drugs from the lower chamber of a double-chamber cell culture system. The applications of the processed calcium phosphate coating was first demonstrated as a drug screening device by using alendronate, a widely used drug for osteoporosis. It was confirmed that the delivery of alendronate significantly decreased both the number of monocyte-differentiated osteoclasts and coating resorption. To demonstrate the application in studying bone metastasis, we delivered PC3 prostate cancer conditioned medium and confirmed that both the differentiation of monocytes into osteoclasts and the osteoclastic resorption of the calcium phosphate coating were significantly enhanced. This novel assay thus provides a new platform for studying osteoclastic activities and assessing drug efficacy *in vitro*.

---

### Introduction

Bones in our skeletal system undergo a continuous remodeling process, in which new bone is formed by osteoblasts and old bone is resorbed by osteoclast[1–3]. Various diseases could break the balanced activity between the two types of cells, and detrimentally affect structural integrity, mobility, and health. Increased osteoclastic activity with ageing tilts the balance towards bone resorption, and leads to a decrease in bone mineral density and eventually osteoporosis, a disease that causes about 9 million bone fractures globally each year[4,5]. Cancer metastasis also disrupt bone homeostasis with devastating consequences. Advanced stages of cancer would metastasize to bone in 65%-75% of breast and prostate cancer[6–8]. Invading cancer cells activate osteoclasts leading to a sustained vicious cycle of bone matrix remodeling and tumor growth[9–11]. Osteoclastic resorption of bone thus plays a central role in both osteoporosis and bone metastasis, and has been the focus of many *in vivo* and *in vitro* studies with the ultimate goal of developing effective therapeutic solutions for the diseases[5,12].

Developing a reliable *in vitro* assay that simulates osteoclastic resorption of bone would be a first step to take before proceeding to any animal or human studies[13–15]. Traditionally, slices made from bone and dentin have been well-accepted for *in vitro* studies on osteoclastic resorption[13,16,17]. Both the number of osteoclasts and the area of the resorption pits can be measured and used as indicators for osteoclast activities[13,16,18–20]. However, the organization of structural elements including porosity and mineral content in dentin and cortical bone vary between species and even within different locations on the same sample, preventing to obtain probes with consistent quality[13].

Synthetic calcium phosphate coatings are a logic alternative to use *in vitro* for osteoclastic resorption activity due to their chemical similarity to bone mineral[14,21–27]. For assaying purpose, a thin calcium phosphate coating on a transparent substrate such as a polymeric disc or the cell culture plate would be a preferred choice over a bulk material since it enables fast and direct assessment of osteoclasts and a reliable measurement of resorption pits under an optical microscope. Such a coating has been processed by electrostatic spray deposition[26], thermal spray technique[27], and the well known, solution-based biomimetic technique[21–25] that has been used to observe and quantify osteoclastic resorption. Corning® Osteo Assay is a commercial product, that has a thin calcium phosphate coating on the bottom well of a polystyrene cell culture plate[28,29]. This assay has been tested for osteoclastic differentiation and growth under the effect of drugs[30–33]. However, those calcium phosphate coating systems, supported in a flat polymeric substrate are not mechanically robust and experience cracking and delamination during sample preparation processes for microscopy, which complicates analysis on resorption pits[28]. More importantly, currently reported assays lack the flexibility of delivering drugs.

Inspired by recent progress in biomineralization studies by Meldrum et al.[34–41], we developed a robust and novel structured calcium phosphate coating with nano-pins on the track-etched porous membrane for the first time. This was achieved through a novel mineralization technique, ammonia-induced mineralization of calcium phosphate (AiM-CaP). The calcium phosphate-coated porous membrane forms the bottom of a cup-shaped cell culture insert. The track-etched pores in the PET membrane allow calcium phosphate mineral pins to grow inside, which, on one hand, enhances coating integration with membrane substrate, and on the other hand provides diffusion channels between inserts and outside culture chamber for biomolecule or drug delivery. This novel assay thus provides a new platform for assessing osteoclastic activities *in vitro*. We demonstrated the applications of the culture insert with AiM-CaP coating first in drug screening for osteoporosis by using alendronate, a widely used drug for osteoporosis, and second in bone metastasis by confirming the effect of prostate cancer cells on osteoclastic resorption.

**Materials and Methods**

**Ammonia-induced mineralization (AiM) of calcium phosphate on cell culture inserts**

Aiming to develop a robust and uniform calcium phosphate coating on the internal bottom surface of the cup-shaped cell culture inserts. We used Falcon® transwell Cell Culture Inserts with pore sizes of 0.4 μm (12 well, high density, translucent polyethylene terephthalate membrane PET, VWR).

The accelerated calcification solution (ACS), a supersaturated calcium phosphate solution[42], was prepared by dissolving 2.32 mM $NH_4H_2PO_4$, 3.87 mM $CaCl_2$, 150 mM NaCl, 40 mM HCl, 50 mM tri(hydroxyme-thyl) aminomethane (Tris) in distilled and deionized $H_2O$ (dd$H_2O$, MilliQ). The pH of the solution was adjusted to 7.4 at room temperature (RT, 20-22°C) by HCl and Tris (All the reagents were Sigma-Aldrich). The ACS solution was stable at 4 °C without precipitation for 2 weeks.

In the typical ammonia-induced precipitation experiment, Falcon® transwell Cell Culture Inserts were filled with ACS solution and sealed by Parafilm (Bemis®). The sealed inserts with ACS were then immersed into ACS solution, the whole solution was degassed under vacuum for 20 minutes to ensure complete filling of the throughout pores with the ACS solution. Each insert was then placed upside down in a 50ml glass beaker, together with a smaller beaker containing freshly made 0.3 ml of 0.5% ammonia solution (Fig. 1). The 50 ml beaker was then sealed and kept at room temperature (25 °C) to allow the diffusion of ammonia gas into the cell culture insert through the porous PET membrane at the base of the insert cup. After 48 hours, the insert was rinsed with ddH2O three times and air-dried.

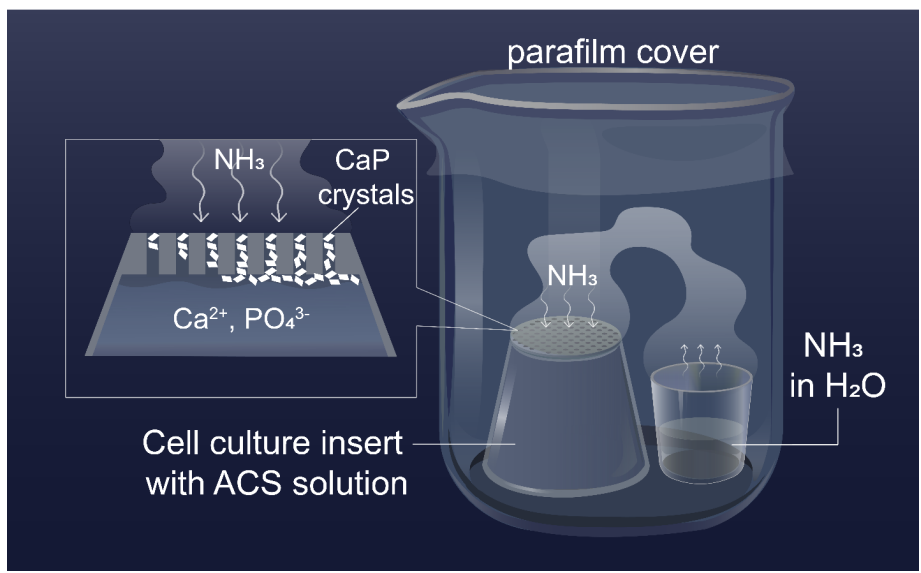

Figure 1. Schematic diagram showing calcium phosphate precipitation in the presence of ammonium hydroxide within track-etch Falcon® cell culture insert. The ACS filled cell culture insert was sealed within a beaker in the presence of ammonium hydroxide solution. The ammonia gas slowly diffused into the insert through the pores, and increased the pH of ACS. The pores and insert are filled with super-statured ACS solution, so precipitation happened in the pores and on the surface of insert membrane.

**Characterization of AiM-CaP coating**

The coating was characterized by scanning electron microscopy (SEM, Hitachi S3000N), X-ray powder diffraction (XRD, Bruker APEX DUO), Fourier-transform infrared spectroscopy (FTIR, PerkinElmer precisely Spectrum 100 FT-IR spectrometer), and inductively coupled plasma-optical emission spectrometry (ICP-OES, Agilent Technologies 5110, USA) to determine the nature of the mineral coating.

To observe surface morphology of the formed CaP coating, the insert membranes were mounted on a SEM stub, sputter-coated with gold/palladium, and observed in a scanning electron microscope equipped with an energy dispersive X-ray spectrometer (SEM/EDS, Hitachi S3000N). An accelerating voltage of 20 kV, 10 mm working distance and high pressure were used for imaging (SE mode). To observe the cross-section morphology, membrane specimens were sputter-coated with carbon. A cross-section was made and imaged by a Focussed Ion Beam-Scanning Electron Microscope (FIB-SEM, Helio NanoLab 650 Focused Ion Beam SEM). An accelerating voltage of 2 kV and high pressure were used for imaging (BSE mode). Triple samples were used in each experiment. The thicknesses of the coating were measured from 10 randomly chosen locations in each sample. The results were presented as means ± standard error.

For chemical composition analysis, the scraped ammonia-induced mineral coating was attached to a glass fiber by oil. The XRD analysis was carried out by an XRD diffractometer (Bruker APEX DUO) with Cu Kα radiation from 2θ 3.5° to 57.46° at a scanning rate of 0.02° s$^{-1}$. The CaP coatings were also analyzed with Fourier-transform infrared spectroscopy (FTIR, PerkinElmer precisely Spectrum 100 FT-IR spectrometer) in transmission mode at the resolution of 1 cm$^{-1}$. Inductively coupled plasma - optical emission spectrometry (ICP-OES, Agilent Technologies 5110) was used to obtain Ca/P ratio of the coatings. The CaP coating on insert membrane was dissolved by immersion in 5 ml of 2% nitric acid solution for 12 hours. The solutions were then filtered using a 0.2 μm filter (Corning sterile syringe filter, Germany) and the Ca and P concentrations in the solution were measured. Then Ca/P ratio were averaged from three specimens. The results were compared with calibration curves made out of standard solutions of Ca and P.

**Osteoclast culture**

Cell culture was done using a dual-chamber setup where the AiM-CaP inserts were placed into a 12-well tissue culture plate (Costar®, Corning). To improve cell adhesion, CaP inserts were incubated in 40 μl of 10 μg/ml vitronectin solution (Sigma-Aldrich) for 12 hours at 4°C. Mouse macrophage-like transformed cells (RAW 264.7, ATCC® TIB-71™) were cultured in DMEM (Dulbecco's Modified Eagle Medium, gibco) with glucose and glutamine, supplemented with 1% penicillin-streptomycin and 10% fetal bovine serum (FBS, Sigma-Aldrich). On day 1, Raw264.7 cells were seeded in each CaP insert at a density of $4 \times 10^3$ cells/cm$^2$ and cultured at 37°C in a humidified atmosphere of 5% CO2. After 24 hours (day 2) of culture, 100 ng/ml RANKL, (Sigma-Aldrich) was added to facilitate the differentiation to osteoclasts. The volume of cell culture medium in the upper chamber (insert) and lower chamber was both 1 ml. Culture medium was changed every 3 days. Osteoclasts and their resorption were evaluated after 7 days.

### The alendronate delivering through AiM-CaP insert

Two different techniques were used to deliver alendronate. For the first delivery technique, alendronate was loaded to the calcium phosphate membrane directly before cell culture (named ALN-CaP insert). ALN-CaP inserts were prepared by soaking CaP inserts in 100 μM alendronate PBS solution for 7 days, following an alendronate loading protocol reported in previous studies[43,44]. For the second delivery technique, after 24 hours of Raw264.7 cell culture, 100 μM alendronate was loaded into the lower chamber of the cell culture system, and allowed to diffuse slowly into the upper chamber through the CaP filled pores (named CaP insert-ALN lower). The ammonia-induced CaP insert was used as control group (named CaP insert). As a negative control, Raw 264.7 cells were cultured on the original track-etched PET membrane, with no added alendronate (named insert). Each group was done in triplicates, and the media was changed every 3 days. At day 8, the cell culture ended, and samples were prepared for analysis. This experiment was repeated three times.

### Osteoclast culture with prostate cancer cell conditioned medium

To demonstrate the applications of our system in studying the influence of prostate cancer cells on osteoclasts, human prostate cancer cell line PC3 (ATCC; Rockville, MD) were cultured in DMEM ($2 \times 10^5$ cells, 150mm petri dish) with glucose and glutamine, supplemented with 1% penicillin-streptomycin and 10% FBS. After 48 h of attachment and proliferation, the medium was replaced with DMEM containing 1% FBS. The PC3 conditioned medium was collected after 48 h, filtered with 0.2 μm filter and stored at -80 °C before use. On the day of using, PC3 conditioned medium (PC3-CM) was prepared by mixing 45% of the stored PC3 conditioned medium and 45% DMEM with 10% FBS[45]. On day 1, Raw 264.7 cells were seeded in each AiM-CaP insert at a density of $2 \times 10^3$ cells/cm$^2$, and then cultured at 37°C in a humidified atmosphere of 5% $CO_2$.

The effects of prostate cancer conditioned medium on osteoclastic activity was studied by comparing the following groups (Table 1). Group a, a control group with presence of RANKL, a key factor for osteoclast differentiation and activation. 100 ng/ml RANKL was added into 90% DMEM-10% FBS after 24 hours of monocyte seeding. The medium in both the insert and the lower chamber was replaced every two days, and cells were cultured in this medium for 7 days (a. RANKL 7). Group b had PC3-CM in CaP inserts and in lower chamber after 24 hours of seeding until the 7[th] day (b. PC3 7). The medium was replaced every two days. According to Alsulaiman et al., PC3-CM only (without addition of RANKL) could induce bone marrow stromal cell into osteoclasts[46]. In order to facilitate osteoclast differentiation, the group c, d and e had 100 ng/ml RANKL in the culture medium both inside the CaP inserts and the lower chamber for the first 4 days. From day 5 to day 7, the RANKL-containing culture medium was replaced by 90% DMEM -10% FBS in group c (c. RANKL+DMEM 4+3) and by PC3-CM in group d (d. RANKL+PC3 4+3) both inside the CaP inserts and the lower chamber. For Group e, only the lower chamber medium was replaced by PC3-CM after the day 4, the medium in the CaP insert was 90% DMEM -10% FBS (e. RANKL+PC3-L 4+3). For group c, d and e, the medium was changed every two days until day 7. Every group had three specimens.

Table 1. Cell culture medium composition in the study of effect of prostate cancer cell's conditioned medium

| Day | a RANKL 7 | | b PC3 7 | | c RANKL+DMEM 4+3 | | d RANKL+PC3 4+3 | | e RANKL+PC3-L 4+3 | |
|---|---|---|---|---|---|---|---|---|---|---|
| | RANKL | PC3 | RANKL | PC3 | RANKL | DMEM | RANKL | PC3 | RANKL | PC3 |
| 1 | + | | | + | + | | + | | + | |
| 2 | + | | | + | + | | + | | + | |
| 3 | + | | | + | + | | + | | + | |
| 4 | + | | | + | + | | + | | + | |
| 5 | + | | | + | | + | | + | | +L |
| 6 | + | | | + | | + | | + | | +L |
| 7 | + | | | + | | + | | + | | +L |

+ : This composition was added into cell culture medium.

**Tartrate-resistant acid phosphatase (TRAP) staining for osteoclasts**

To visualize morphology of osteoclasts and to quantify osteoclast activation, cells were stained for TRAP activity, which is an enzymatic marker of osteoclasts[47]. The cells were fixed in 4% paraformaldehyde-phosphate buffer solution for 15 minutes and rinsed with PBS. An Acid Phosphatase, Leukocyte (TRAP) Kit (387A, Sigma Aldrich) was used following the protocol provided by the manufacture. After 1-hour staining at 37°C, cells were rinsed with PBS and air-dried. Osteoclasts were identified as multinucleated TRAP-positive cells and their number was counted under an optical microscope in triplicate (OM, Zeiss AXIO Zoom. V16 stereo zoom microscope).

## Fluorescence microscopy of osteoclasts

Fluorescence microscopy (Zeiss Axiophot) was used to visualize the TRAP, cytoskeleton and nuclei of osteoclasts. The samples were fixed in 4% paraformaldehyde-phosphate buffer solution for 30 minutes, and then permeabilized with 0.1% Triton-X 100 in PBS for 5 min. Non-specific sites were blocked with 2% BSA in PBS for 30 minutes. The samples were first stained with ELF 97 (Thermo Fisher) The acid phosphatase solution contained 9 ml of pre-warmed ddH$_2$O, 400 μl of acetate solution, 100 μl of naphthol AS-BI phosphoric acid, 200 μl of tartrate solution, and 200 μl of diazotized Fast Garmet GBC. The samples were incubated for 15 minutes at 37 °C in darkness with ELF 97 phosphatase solution, which was freshly prepared by diluting ELF 97 dye 60 times in acid phosphatase solution[48]. Then the samples were then incubated with TRITC-conjugated phalloidin (Millipore-Sigma) (0.5 μg/ml in 2% BSA solution) for 90 minutes at room temperature in the dark. Last, the samples were incubated with DAPI (Millipore-Sigma) (1 μg/ml in 2% BSA solution) for 5 minutes at room temperature in darkness. Between each staining step, samples were rinsed with PBS. The membrane at the base of the inserts was cut off from inserts and mounted with ELF 97 mounting media on a glass slide to preserve fluorescence.

## Osteoclast interaction with AiM-CaP coating

To visualize osteoclasts and their interactions with the calcium phosphate coating, SEM analysis was performed. After 7 days of culture, three specimens from each group were fixed in 4% paraformaldehyde-phosphate buffer solution for 15 minutes, followed by 2.5% glutaraldehyde-PB solution for 20 minutes. The samples were then post-fixed with 1% osmium tetroxide (OsO$_4$)-PB solution for 60 minutes. The samples were dehydrated in ascending concentrations of ethanol (30%, 50%, 70%, 95% and 100%) for 10 minutes each, followed by CO$_2$ critical point drying. To study detailed interaction between osteoclasts and AiM-CaP coatings, cross-sections through the cells and the underneath CaP coating were prepared and imaged by FIB-SEM (Helio NanoLab 650 Focused Ion Beam SEM). SEM observation was performed as previously described. After SEM observation, the samples were coated with 20nm of iridium (Leica EM MED020 Coating System). In order to obtain a smoother cross-section, a thin layer of platinum (Pt) was deposited on the sample surface in order to reduce the curtaining effect. An accelerating voltage of 2 kV and high pressure were used for imaging (BSE mode).

**Osteoclastic resorption activity**

To quantify the areas of the resorption pits by osteoclasts on the calcium phosphate coatings, all the cells on the coating surface were washed three times by $ddH_2O$ on day 7. The cells could be removed due to different ion concentration and osmolality. The surface topography of CaP substrate after incubation with osteoclasts was examined by an SEM (Hitachi S3000N) using back-scattered electron signal to differentiate resorbed areas from the coating as previously described. The elemental distribution was examined by EDS (Hitachi S3000N). The total area of the osteoclastic resorption pits was determined by imaging 25 squares (each 1.2 mm X 1.0 mm) uniformly distributed over each CaP coating surface, followed by ImageJ quantitative analysis. For each group, triple specimens were used for quantification.

**Statistical analyses**

Statistical analysis was performed by SPSS 25.0 software (SPSS Inc, Chicago, IL, USA). The number of osteoclasts and area of resorption pits data are presented as means ± standard error. Differences between groups were analyzed by Mann-Whitney U test. Statistical significance between two groups was established at the P value lower than 0.05. All experiments were carried out in triplicate.

**Results**

**Homogeneous nano-structured calcium phosphate coating generated by ammonia-induced mineralization**

The original track-etched PET membrane covering the bottom of the cell culture inserts have a high density of throughout pores ($1\times10^8$ pores/cm$^2$) of 400 nm diameter (Fig. 2a). After 48 hours of ammonia-induced mineralization (AiM), the cross-sectioning through the membrane by focused ion beam revealed the unique structure of the mineral coating, which had a uniform thickness with the complete filling pores of minerals (Fig. 2b). The inner surface of the PET membrane, which is the surface exposed to the accelerated calcification solution (ACS), was fully covered by flake-shaped minerals (Fig. 2c). On the outer surface of the membrane (i.e. the exterior surface of the cell culture inert), which was exposed to ammonia gas, we observed that only the nano-pores were filled with the minerals (Fig. 2d). Such an ammonia-induced mineralization (AiM) process (Fig. 1), generates flake-like minerals that first fill the nano-pores, and then formed a uniform coating on the inner surface of the PET membrane. The thickness of the mineral coating was $246\pm25$ nm (Fig. 2b).

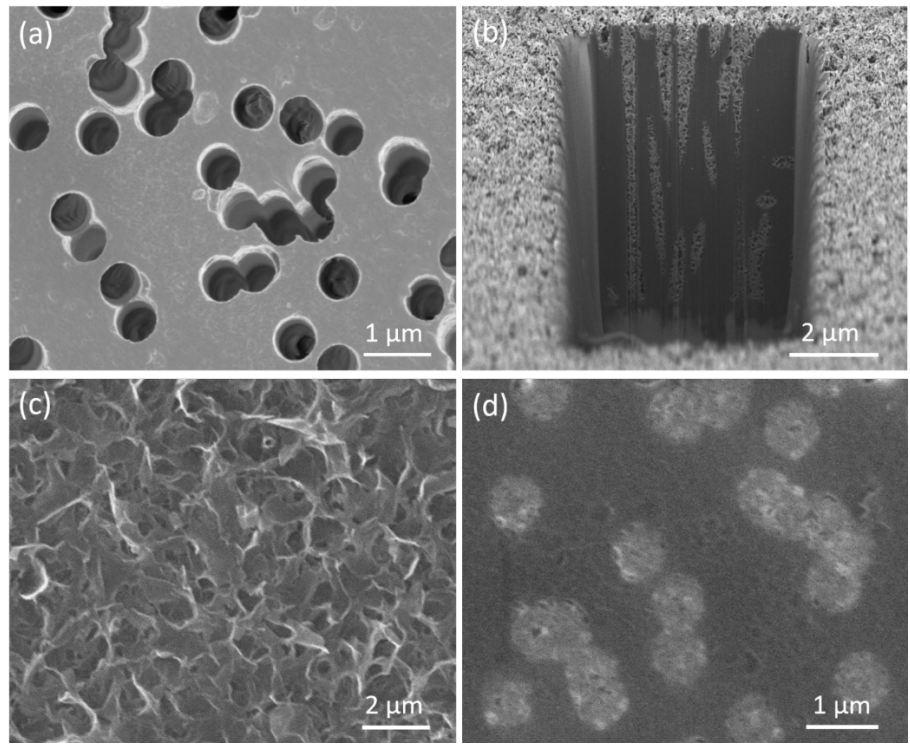

**Figure 2.** SEM images of the original track-etched Falcon® cell culture insert membrane showing nano-pores of 400nm in diameter (a); focused ion beam cross section showing calcium phosphate minerals forming coating on membrane surface and filling the throughout pores (b); (c) calcium phosphate coating on the inner surface of insert; (d) outer surface of the insert showing calcium phosphate (bright phase) only filling the pores.

XRD spectrum showed the major diffraction peaks of octacalcium phosphate (OCP) at $2\theta$ 4.7°, 9.7°, and 16.0°[49,50] (Fig. 3a). Since HA and OCP have many overlapping peaks, HA could not be excluded by XRD spectrum[49,50]. The FTIR spectrum of AiM-CaP coating showed peaks at 472 cm[-1], 563 cm[-1], and 601 cm[-1] (Fig. 3b). According to the literature, the peaks at 472 cm[-1] corresponds to the $U_{2a}$ absorption O-P-O bands of the HA. The peaks at 561 and 602 cm[-1] are $U_{4c}$ and $U_{4a}$ absorption O-P-O bands of the HA respectively (Fig. 4.8b)[51,52]. The OCP has two peaks at 560 and 603 cm[-1], but it doesn't has peak around 472 cm[-1][52–54]. Comparing with published results, the spectra of ammonia-induced mineralization coating was closer to HA or more precisely carbonated apatite since hydroxyl signal was absent. The ICP-OES results showed the Ca/P ratio to be 1.55±0.04, which is between that of octacalcium phosphate (OCP, 1.33) and stoichiometric hydroxyapatite (HA, 1.67). Therefore, the chemical composition of AiM-CaP coating was a mixture of OCP and carbonated apatite. The final structure is thus a thin CaP coating robustly anchored to the porous PET membrane with numerous nano-sized CaP pins.

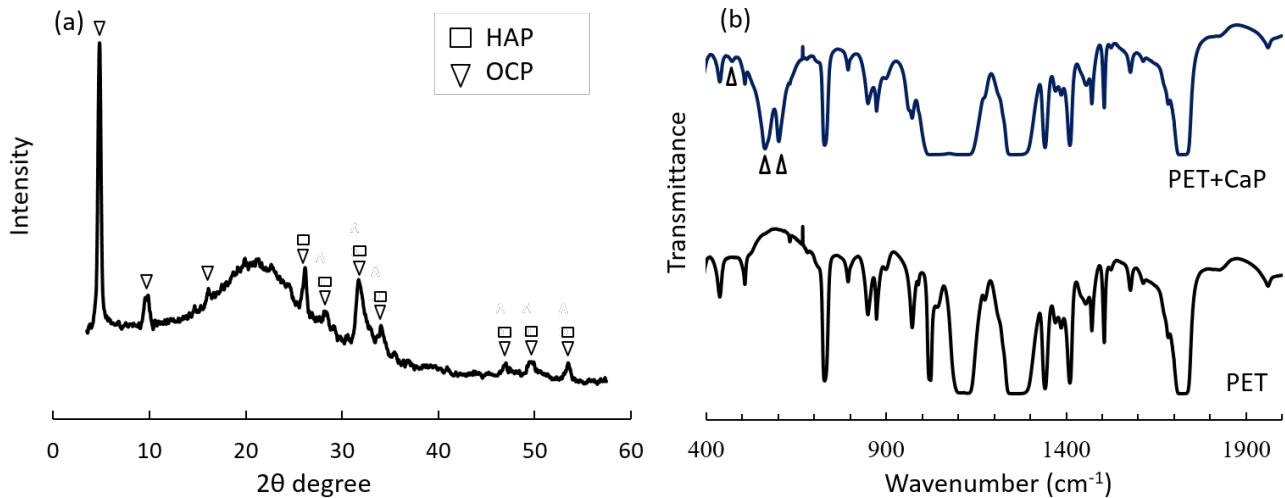

**Figure 3. Chemical composition of ammonia-induced CaP coating.** (a) XRD spectrum of the calcium phosphate coating on the inner surface of insert; (b) FTIR spectra of the calcium phosphate coating on insert membrane surface (PET+CaP) and original insert membrane (PET). Triangles mark CaP peaks.

**Osteoclast activity evaluation and imaging on AiM-CaP coating**

In order to observe the usefulness of the AiM-CaP coating for cell culture, we evaluate Raw 264.7 differentiation into osteoclasts. RAW 267.4 is a mouse leukemic monocyte macrophage cell line commonly used in osteoclast studies. The RAW264.7 cells could generate large TRAP[+] cells capable of bone resorption in the presence of RANKL without M-CSF[13,55]. Comparing with human primary osteoclasts, the RAW 264.7 cells are widely available and easy to culture; they have a stable nature of the precursor populations, and are more amenable to genetic manipulation[56,57]. Therefore, Raw 264.7 were used to in this osteoclastic resorption study.

After 7 days of culture in presence of RANKL, RAW 264.7 cells were able to differentiate into osteoclasts on the AiM-CaP coating. They appeared purple after TRAP staining (Figure 4a), which is a well-established histochemical marker for osteoclasts [47]. The cytoskeletal actin ring on multinucleated osteoclasts, which is indicative of mature cells, could be seen under a fluorescence microscope after the specimens were stained with TRITC-conjugated phalloidin (Fig. 4b, arrow). These results shows that osteoclasts could be differentiated from monocytes on AiM-CaP coatings.

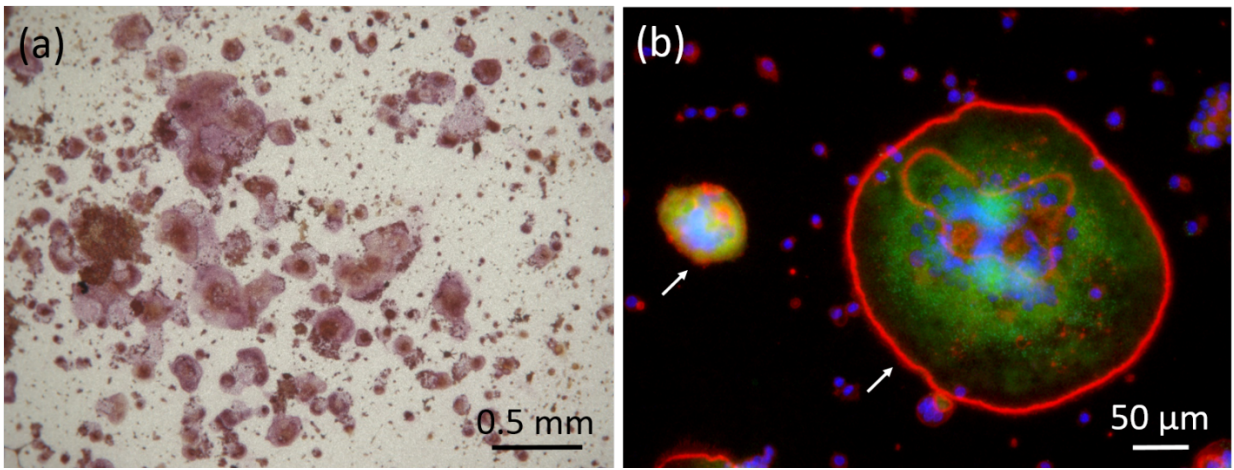

**Figure 4. Osteoclasts on ammonia-induced calcium phosphate insert.** (a) Tartrate resistant acid phosphatase (TRAP) staining. (b) Immunofluorescence image of osteoclast (white arrow) on coating surface. Cells were fixed, labeled for TRAP (green, ELF97), nuclei (blue, DAPI) and F-actin (red, TRITC).

FIB-SEM images in Figure 5 show an active osteoclast and its interaction with the AiM-CaP coating. A top view of the osteoclast shows the CaP coating to its right side has been partially resorbed, leaving a shallow pit behind and revealing the mobile nature of the cell (Fig. 5a, 5b). The cell also resorbed both the CaP coating and the mineral pins deep inside PET membrane on upper boarder, exposing the original through pores in the membrane. A cross-section of the osteoclast made by focused ion beam inside the SEM showed that the CaP coating underneath the cell body has been completely resorbed (Fig. 5c, 5d), and the osteoclast is in close contact with the PET membrane. The CaP minerals inside the nano-pores have been partially resorbed, and empty pores are now occupied by presumably cell processes (Fig. 5d, 5e). As comparison, areas without osteoclasts shows a thin layer of CaP coating and fully filled CaP minerals (Fig. 5f).

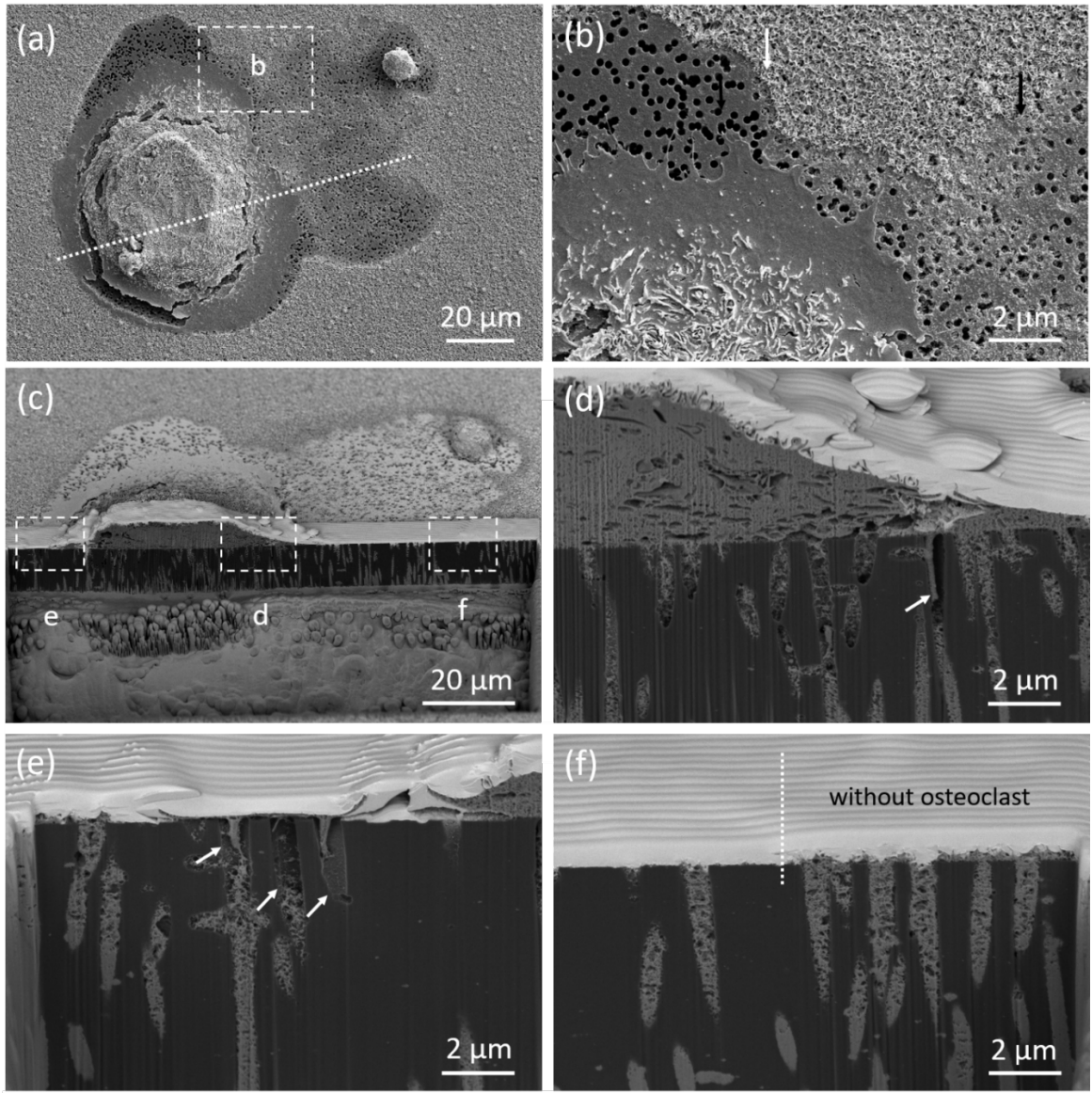

**Figure 5. SEM and FIB images of osteoclast interaction on calcium phosphate coating.** (a) Top view of osteoclast and its resorption pit on ammonia-induced calcium phosphate coating. Focused ion beam cut the cells and membrane along the white dot line. (b) Magnified resorption areas of osteoclast. The white arrow indicated deep resorption pit, and the black arrow indicated the shallow resorption pit. (c)

The cross section view of osteoclast on calcium phosphate coating. Underneath the osteoclast (d) and at the edge of the osteoclast (e), there were partially resorbed pores indicated by white arrow. (f) At the shallow resorption pit area, the coating on the surface was resorbed, but the minerals within the throughout pores remained.

When osteoclasts have been removed, the resorption pits can be easily seen under SEM using backscattered electron imaging (Fig. 6). After resorption of CaP coating, the PET polymer is exposed, which generates weaker BSE signal than the CaP coating. Such a high contrast between the resorption pits and the surrounding CaP coating makes it possible to quantify osteoclast activity. The AiM-CaP coating on track-etched PET membrane thus provides the potential to study osteoclast resorption of calcium phosphate minerals

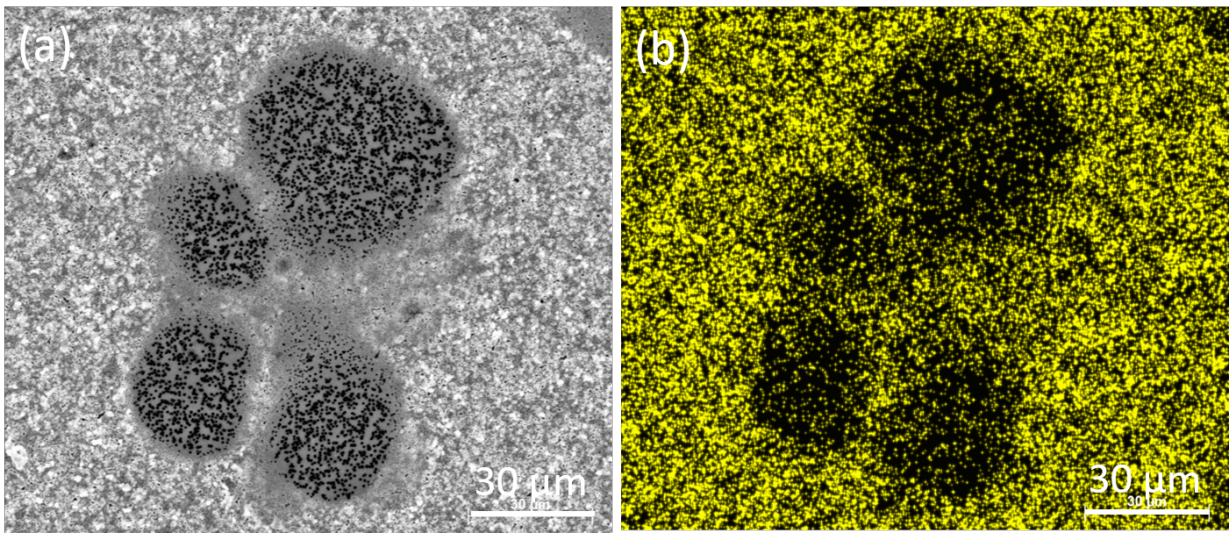

**Figure 6. Resorption pits on ammonia-induced calcium phosphate coating surface.** (a) BSE image of resorption pits (dark area) and calcium phosphate coating (bright area). (b) EDS mapping of calcium.

**Alendronate inhibition on osteoclastic resorption through AiM-CaP coating**

We first demonstrated the application the AiM-CaP coated culture inserts by testing the effect of alendronate on osteoclastic resorption. Alendronate is the most widely used osteoclast-inhibiting bisphosphonate drug for treating osteoporosis and Paget's disease of bone. It is thus expected to inhibit the osteoclastic resorption of the CaP coating. In this study, alendronate was delivered in two ways: by direct absorption onto the CaP coating before cell culture, and by diffusing into the porous CaP nano-pins through the lower cell culture chamber during cell growth. Figures 7 a-c show representative morphologies of the TRAP-stained osteoclasts in each group, and figure 7 d-f are representative BSE images of the resorption pits. Without alendronate drug (CaP insert group), osteoclasts were active on AiM-CaP surface and created deep resorption pits (Fig.7a and d), with 880±110 osteoclasts per CaP insert and areal resorption percentage of 17.7±3.3 % (Fig. 7 g and h).

On AiM-CaP surface pre-loaded with alendronate (ALN-CaP insert group), both the number of osteoclasts (47±20) and the areal resorption percentage (0.6±0.6 %) dramatically and significantly decreased (Figs. 7 b, e, g and h). In the other testing group when alendronate diffused from lower chamber through the pores during cell culture (CaP insert-ALN lower group), the osteoclastic activity was also significantly inhibited (Fig. 7c, f, g and h) as compared with the control group with 288±89 osteoclasts and an areal resorption of 1.56±0.7 %. The resorption pits were also notably shallower than the control group (CaP insert group). There is no significant difference in osteoclast activity between the two alendronate delivery methods. As a comparison, there were 115±90 osteoclasts on the original insert (insert group), which was without CaP coating and ALN (Fig. 7g). These results demonstrated that the AiM-CaP coated cell culture inserts could be used to quantitatively assess osteoclastic resorption activity and to test drug efficacy.

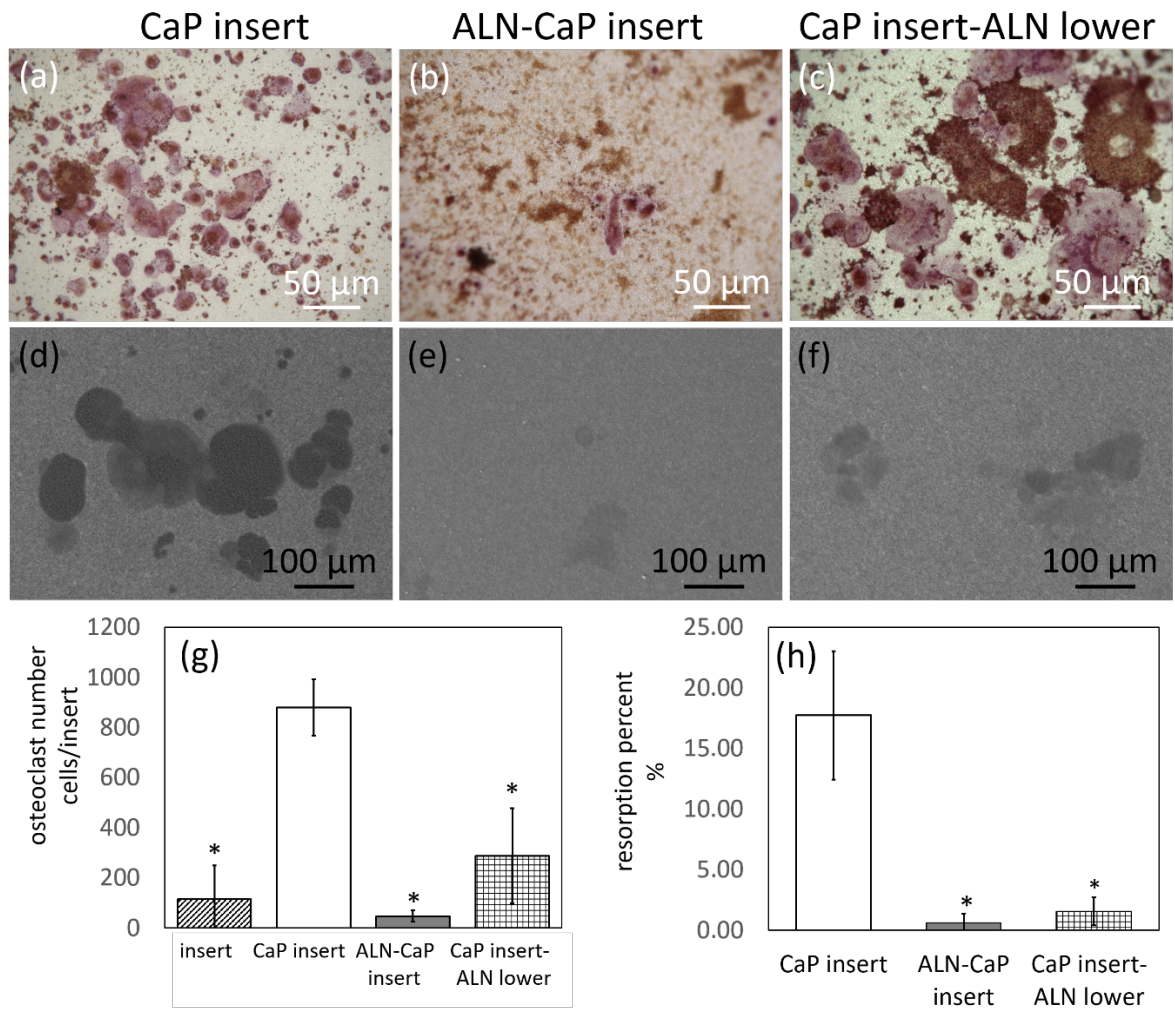

**Figure 7.** Representative optical microscopy (OM) images of TRAP positive osteoclasts (a-c) and BSE images of osteoclastic resorption pits (d-f) on the surface of CaP inserts (a and d), ALN-CaP inserts (b and e), and CaP insert-ALN (c and f). (g) The number of TRAP positive osteoclasts on different surface. (h) The resorption area ratio on coating surface. * Indicates statistically significant difference (P < 0.05) compared to CaP insert group, which was control group.

**Effect of prostate cancer cell conditioned medium on osteoclastic resorption activity**

To explore the feasibility of applying the AiM-CaP coating for studying the osteoclastic effects of bone metastasis, or other stimulatory conditions that could enhance osteoclasts activation and bone resorption activity, we used prostate cancer cells (PC3) as a model to stimulate osteoclast proliferation and resorption. For that purpose, the conditioned media obtained from PC3 cells was used (osteolytic prostate cancer cell line) to stimulate RAW 264.7-derived osteoclasts. We performed two methods of loading the conditioned media, one in the cell culture medium (directly on the osteoclasts) and one in the lower chamber which would depend on diffusion through the AiM-CaP coated membrane. Figure 8 a-e and f-j showed representative morphologies of the osteoclasts and the BSE images of their resorption pits. In PC3 7 group without RANKL, there were no osteoclasts formed and no resorption observed (Fig. 8 b and g). When adding RANKL in the first 4 days of cell culture, followed by adding PC3-CM for three days, osteoclasts formed on CaP coating, although their size was smaller than those cells treated with RANKL for 7 days (Fig. 8 d, i and a, f). When PC3-CM was added into the lower chamber, osteoclasts and their resorption pits were smaller than RANKL+PC3-CM 4+3 group (Fig. 8 e, j and d, i).

A comparison between PC3 7 group and RANKL 7 group showed that PC3 conditioned medium alone could not enhance osteoclast differentiation and resorption activity (0±0) (Fig. 8 b, g and a, f). However, by comparing RANKL+PC3 4+3 with RANKL+DMEM 4+3 groups (Fig.8 d, i and c, h), we can conclude that adding PC3 conditioned medium after 4 days RANKL treatment (RANKL+PC3 4+3), both osteoclast numbers (142±26) and resorption activity (9.6±0.6%) increased significantly, from 31±5 and 0±0 % for the RANKL+DMEM 4+3 group. When PC3-CM was added into lower chamber of CaP inserts (RANKL+PC3-L 4+3), and diffused to the culture inserts through the porous CaP nanopins, both the osteoclast number (87±10) and resorption area (5.7±0.7%) were 39% and 41% lower than the RANKL+PC3 4+3 group, respectively (Fig. 8 e, j and d, i). As a comparison, the RANKL 7 group has the osteoclast numbers of 330±15, the areal resorption percentage of 11.5±0.8 % (Fig. 8 a and f). These results confirms that RANKL is necessary for initiating formation of osteoclasts, and PC3 conditioned medium could further stimulate osteoclast proliferation and resorption activity. Importantly, the stimulation that we demonstrated by lower chamber PC3-CM loading, suggests that AiM-CaP inserts have an enormous potential to be used in co-culture systems to study cell interaction of osteoblasts/osteoclasts lineages with other cell types as it could be the metastases of bone homing cancers.

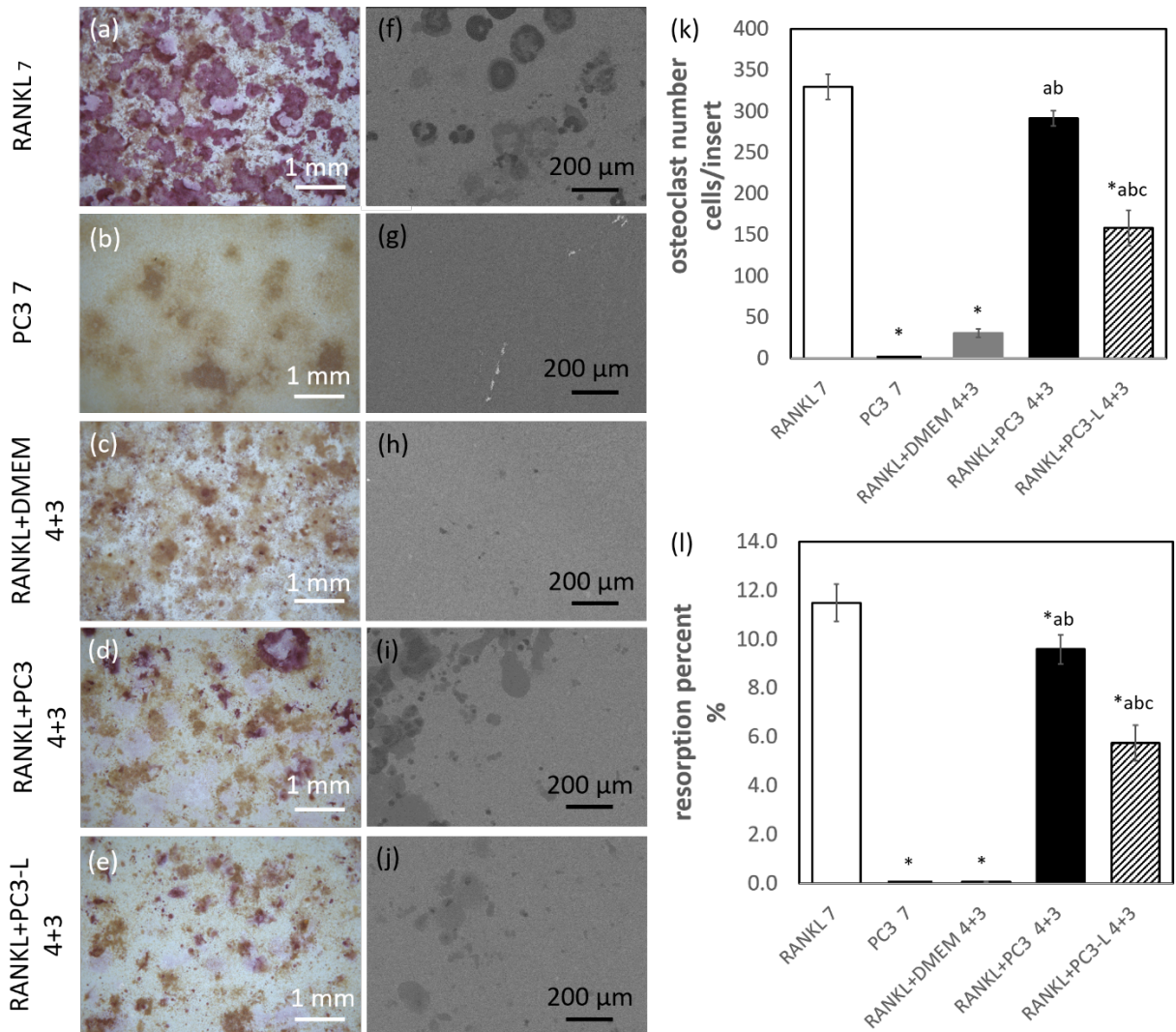

**Figure 8.** Representative OM images of TRAP positive osteoclasts (a-e) and BSE images of osteoclastic resorption pits (f-j) on the surface of CaP inserts. (k) The number of TRAP positive osteoclasts. (l)The resorption area ratio on coating surface. * Indicates statistically significant difference (P < 0.05) compared to RANKL 7 group, which was control group. Statistically significant differences between groups were indicated with letters (a, b, c) where a, b and c indicates significant difference (P < 0.05) compared to PC3 7, RANKL+DMEM 4+3, and RANKL+PC3 4+3, respectively.

**Discussion**

The calcium phosphate coating generated by ammonia-induced mineralization (AiM) on track-etched PET membrane has unique nano-structure and functions. The coating is thin and uniform, and only covers one side of the membrane for testing osteoclastic resorption. The calcium phosphate nano-rods serve a dual function: they mechanically anchor the coating to the PET membrane thus preventing coating spall-ation, while the porous nature of the calcium phosphate nano-rods makes it possible to deliver drugs or conditioned medium from the lower chamber of a double-chamber cell culture system. This study demonstrated that such a device can be used as a platform to study osteoclasts. The results of alendronate delivery and cancer cell conditioned medium delivery showed the potential of using this platform for new drug testing and co-culture in the future.

Making a nano-structured calcium phosphate coating was originally inspired by the biomineralization studies on track-etched polymer membranes[34,35,38,40,58,59]. Meldrum and co-workers used a double-diffusion method and investigated the formation of calcium carbonate[34,35,38,58], calcium oxalate[59] and calcium phosphate[36,40] minerals within the nano-pores of the track-etch membranes. In their studies, a track-etched membrane was sealed between a pair of U-tube arms that contained $Ca^{2+}$ solution in one arm and anion (e.g. $PO_4^{3-}$) solutions in the other arm. The cations and the anions diffused towards each other inside the nano-pores and reacted to form mineral rods. When applying the double diffusion method to our study, we found it was not possible to obtain a uniform and thin (less than a micron) coating on only one surface of the membrane, and instead a thick coating formed on both sides of the membrane (Supporting Information Fig. S5). We thus developed the ammonia-induced mineralization (AiM) technique, as illustrated in Figure 1. The influence of the ammonia concentration, the mineralization time, the throughout pore size and the different mineralization solutions on the resulting minerals was investigated (Supporting Information Fig. S1-4 and Table S1). The optimized influence parameters were used for AiM-CaP insert here. In a sealed beaker, ammonia gas diffuses into the transwell insert through the track-etched pores and increases the pH of the calcium phosphate solution inside the insert cup. As a result of the pH increase, the solution becomes supersaturated and starts to precipitate calcium phosphate minerals[60]. The unidirectional diffusion also creates a gradient in pH, which is the highest at the entry point of the nano-pores (i.e. at the exterior surface of the cell culture inert) and decreases towards the inner side of the cup. Therefore, mineralization starts inside the nano-pores at the exterior surface of the cell culture insert and grows along the nano-pores following the diffusion direction of the ammonia gas (Figures 1, 2d and Supporting Information Fig. S2). When the mineralization process reaches the interior surface of the culture insert, CaP crystals start to grow laterally along the surface of the PET membrane that acts as the heterogeneous nucleation site. As a result, a thin coating of CaP is formed at the bottom of the culture insert. As the CaP crystals grow inside the nano-pores, they progressively slow down the diffusion of ammonia and thus the increase of pH. Therefore, CaP crystal growth is stable and easy to control. In our study, the flake-like CaP minerals could be seen in the nano-pores under SEM six hours after ammonia introduction. A CaP coating started to form at 12 hours (~0.13 μm), and reached to the final thickness of 0.25 μm at 48 hours (Supporting Information Table S1). Such a thin coating is preferred over thicker coating for assessing osteoclastic activities. Upon resorption of the thin CaP coating, the underneath PET substrate is exposed, which creates a high contrast under SEM backscattered electron imaging and enables easy quantification of osteoclastic resorption. The AiM technique reported here is different from the classical ammonia carbonate technique used in the biomineralization study of calcium carbonate or calcium phosphate[35,61–63]. In those studies, ammonium carbonate powder was placed next to a $Ca^{2+}$ containing solution. The decomposition of ammonium carbonate generates ammonia and carbon dioxide gases that would dissolve into the $Ca^{2+}$ solution. Carbonate ions would react calcium ions and precipitate minerals, while the dissolution of ammonia also increases solution pH and further enhances degree of supersaturation.  However, for calcium phosphate precipitation, ammonium carbonate is too mild and cannot easily achieve the solution conditions for CaP precipitation. In the report by Tovani et al., poly(acrylic acid) has to be used to concentrate calcium and phosphate ions into the nanopores in the track-etched membrane[62]. Even with such an effort, the nano-channels could not be fully filled with CaP minerals because of limited supply of ions. We also tried the ammonium carbonate method in our pilot experiments. The method failed to create a continuous coating in the cell insert using ACS solution other than isolated mineral particles (Supporting Information Fig. S6).

The currently available *in vitro* osteoclastic resorption platform has various limitations, which would cause trouble in resorption quantification. In this study, we developed ammonia-induced CaP insert as *in vitro* platform for osteoclastic resorption study. Osteoclastic differentiation and resorption activities were evaluated by number of osteoclast and area of resorption pits respectively. The detailed interaction between osteoclast and CaP coating was revealed by FIB. The enhancing and inhibiting of osteoclastic resorption activity were demonstrated by delivering of alendronate drug and prostate cancer cell conditioned medium culturing through this platform. This ammonia-induced CaP insert provides a stable and novel *in vitro* platform for osteoclast study. The results of alendronate delivery and cancer cell conditioned medium delivery showed the potential of using this platform for new drug testing and co-culture in the future.

Comparing AiM-CaP coating with bone slice in terms of resorption in vitro, the osteoclasts left shallow resorption pits on AiM-CaP insert, but they left irregular shape of resorption pits with various depth on bone slice. On the bone slice, the depth of resorption pits varied from 2 μm to 70 μm[16-20]. In order to quantify the resorption activity accurately on bone slice, a three dimensional measurement method was necessary, and it increased the difficulties of quantification[13]. Additionally, cell lacunae in bone may also cause difficulties in differentiating the newly formed resorption pits from the pre-existing cavities[13]. On the AiM-CaP coating, the resorption depth was controlled to the coating thickness (about 250 nm). In this case, the quantification of resorption activity could be simplified as counting resorption areas on the AiM-CaP coating surface.

Compared with a thin CaP coating on a solid substrate, our AiM-CaP coating offers a few advantages. First, the AiM-CaP coating is mechanically robust due to the CaP nanorods anchored into the track-etched PET membrane. Such a design not only enables the coating to survive osteoclast culture process (Supporting Information Fig.S7 and Fig. S8), but also enables the coating to survive various post cell culture steps without cracking and spallation (e.g. rinsing, drying, and vacuum) before and during SEM imaging. Such a property makes it feasible for using SEM to quantify re-sorption area and to study osteoclast-surface interaction. Commercial products such as Corning® Osteo Assay are more delicate and would require careful storage in water to avoid cracking of the dry mineral surface[30,33]. Second, the presence of the porous CaP nanorods across the track-etched PET membrane enables more flexible delivery routes for drugs and other molecules. In addition to the direct loading to the CaP coating before cell culture, drugs can now be delivered to the coatings and the cells at any designated time through the lower chamber of a double-chamber co-culture system to better study their effects on different stages of cell differentiation and proliferation. This feature is clearly demonstrated in the delivery of alendronate and PC3 conditioned medium. Third, the individual AiM-CaP insert is convenient to perform different staining and characteristic analysis. Quantifying resorption pits, fluorescence staining and TRAP staining require different procedure, such as: fixation process, incubator temperature, and staining time. When the cells in the same plate, it is hard to perform all these staining and analysis together, unless breaking the plate.

Alendronate is one of the bisphosphonates that are known for their pharmacological effect against osteoclast-mediated bone loss due to diseases such as osteoporosis and bone metastasis[64–66]. In this study, the antiresorptive effect of alendronate on osteoclasts is clearly confirmed. Furthermore, the delivery of alendronate through lower cell culture chamber is as effective as the direct loading onto the CaP coating. The alendronate concentration effect on Raw 264.7 cells proliferation was presented in the Supporting Information (Supporting Information Fig. S9). This feature makes future drug screening using the processed device more feasible.

In an attempt to apply the developed device to studying cell-cell communication, we choose bone metastasis in which the interaction between the cancer cells and bone cells (both osteoblasts and osteoclasts) plays a critical role[6]. The cancer cell line used in this study (PC3) is a human prostate cancer cell line frequently used in prostate cancer research and drug development[67]. PC3 cells have been shown to induce differentiation of monocytes (RAW 264.7) both through direct cell-cell contact and the secreted exogenous factors[68]. As a first step, PC3 cell conditioned medium was loaded to the lower chamber of the double-chamber culture system. Our results showed that PC3 cell conditioned medium stimulated monocyte differentiation into osteoclasts and their resorption of the calcium phosphate coating.

**Conclusions**

For the first time, we processed a unique and novel nano-structured calcium phosphate coating by ammonia-induced mineralization (AiM) method on track-etched PET membrane. Through osteoclast culture, this coating was demonstrated to be an excellent candidate for osteoclastic resorption assay *in vitro*. The alendronate experiment proved that AiM-CaP coating could be used for drug delivery and anti-osteoporosis drug test. The prostate cancer conditioned medium experiment proved that cancer conditioned medium would increase osteoclastic resorption activity.

In our study, AiM-CaP coating was generated with transwell insert. The porous flake-like OCP/HA coating had a uniform thickness around 250nm. It had the unique CaP pins within track-etched throughout pores. The TRAP staining, fluorescence staining, and resorption pits demonstrated that this coating is suitable for osteoclast culture and activity evaluation. We also demonstrated the application of AiM-CaP coated insert in drug delivery and test. Alendronate, as an osteoclastic resorption inhibitor, decreased osteoclast resorption through two different paths (direct loading and diffusion through lower chamber). The PC3 conditioned medium, used as osteoclastic resorption stimulator, increased osteoclastic resorption when added directly into cell culture medium and delivered through the lower chamber. This study demonstrated that AiM-CaP coating could be used for cancer cell induced osteoporosis study *in vitro*.

In conclusion, AiM-CaP coating is a unique and reliable platform to study osteoclasts. This platform also provides wide opportunities for coating modification, material-bone cell interaction, new anti-osteoporosis drug test and bone cell metastases study in the future.

## SUPPORTING INFORMATION

The Supporting Information is available free of charge on the ACS Publications website.

Supplementary methods, results, supporting figures, and supporting tables (PDF)

## AUTHOR INFORMATION

### Corresponding Author

**Rizhi Wang**-Department of Materials Engineering University of British Columbia 309-6350 Stores Road, Vancouver, BC V6T 1Z4 Canada. Email: rizhi.wang@ubc.ca

### Authors

**Sijia Chen-**Department of Materials Engineering, University of British Columbia, 309-6350 Stores Road, Vancouver, BC, V6T 1Z4 Canada, Centre for Hip Health and Mobility, University of British Columbia, 2635 Laurel St, Vancouver, BC V5Z 1M9, Canada

**Qiong Wang-** Department of Materials Engineering, University of British Columbia, 309-6350 Stores Road, Vancouver, BC, V6T 1Z4 Canada, School of Biomedical Engineering, University of British Columbia, 2222 Health Sciences Mall Biomedical Research Centre (BRC),Vancouver, BC V6T 1Z3, Canada, Centre for Hip Health and Mobility, University of British Columbia, 2635 Laurel St, Vancouver, BC V5Z 1M9, Canada

**Felipe Eltit-**Department of Materials Engineering, University of British Columbia, 309-6350 Stores Road, Vancouver, BC, V6T 1Z4 Canada, School of Biomedical Engineering, University of British Columbia, 2222 Health Sciences Mall Biomedical Research Centre (BRC),Vancouver, BC V6T 1Z3, Canada, Centre for Hip Health and Mobility, University of British Columbia, 2635 Laurel St, Vancouver, BC V5Z 1M9, Canada

**Yubin Guo-**Vancouver Prostate Centre, University of British Columbia, 2660 Oak St, Vancouver, BC, V6H 3Z6, Canada.

**Michael Cox-** Vancouver Prostate Centre, University of British Columbia, 2660 Oak St, Vancouver, BC, V6H 3Z6, Canada.

**Author Contributions**

Rizhi Wang: Conceptualization, Methodology, Formal analysis, Supervision, Writing-Reviewing and Editing

## ACKNOWLEDGMENT

The research was supported by Natural Sciences and Engineering Research Council of Canada (NSERC Discovery), and the Collaborative Health Research Projects (CHRP) jointly funded by NSERC and the Canadian Institutes of Health Research. Sijia Chen was supported by the China Scholarship Council (CSC) from the Ministry of Education of P.R. China. The authors would like to thank Noa Wang for her contribution to the graphic art (Figure 1).

## REFERENCES

(1)    Robling, A. G.; Castillo, A. B.; Turner, C. H. Biomechanical and Molecular Regulation of Bone Remodeling. *Annu. Rev. Biomed. Eng.* **2006**, *8* (1), 455–498. https://doi.org/10.1146/annurev.bioeng.8.061505.095721.

(2)    Langdahl, B.; Ferrari, S.; Dempster, D. W. Bone Modeling and Remodeling: Potential as Therapeutic Targets for the Treatment of Osteoporosis. *Ther. Adv. Musculoskelet. Dis.* **2016**, *8* (6), 225–235. https://doi.org/10.1177/1759720X16670154.

(3)    Sobacchi, C.; Schulz, A.; Coxon, F. P.; Villa, A.; Helfrich, M. H. Osteopetrosis: Genetics, Treatment and New Insights into Osteoclast Function. *Nat. Rev. Endocrinol.* **2013**, *9* (9), 522–536. https://doi.org/10.1038/nrendo.2013.137.

(4)    Briggs, A. M.; Cross, M. J.; Hoy, D. G.; Sànchez-Riera, L.; Blyth, F. M.; Woolf, A. D.; March, L. Musculoskeletal Health Conditions Represent a Global Threat to Healthy Aging: A Report for the 2015 World Health Organization World Report on Ageing and Health. *Gerontologist* **2016**, *56*, S243–S255. https://doi.org/10.1093/geront/gnw002.

(5)    Rachner, T. D.; Khosla, S.; Hofbauer, L. C. Osteoporosis: Now and the Future. *Lancet* **2011**, *377* (9773), 1276–1287. https://doi.org/10.1016/S0140-6736(10)62349-5.

(6)    Croucher, P. I.; McDonald, M. M.; Martin, T. J. Bone Metastasis: The Importance of the Neighbourhood. *Nat. Rev. Cancer* **2016**, *16* (6), 373–386. https://doi.org/10.1038/nrc.2016.44.

(7)    E. Fenig*, M. Mishaeli*, Y. K. and M. L. Metastatic Bone Disease: Clinical Features, Pathophysiology and Treatment Strategies. *Cancer Treat. Rev.* **2000**, *27* (4), 269–286. https://doi.org/10.1053/ctr.

(8)    Roodman, G. D. Biology of Osteoclast Activation in Cancer. *J. Clin. Oncol.* **2001**, *19* (15), 3562–3571. https://doi.org/10.1200/JCO.2001.19.15.3562.

(9)    Sturge, J.; Caley, M. P.; Waxman, J. Bone Metastasis in Prostate Cancer: Emerging Therapeutic Strategies. *Nat. Rev. Clin. Oncol.* **2011**, *8* (6), 357–368. https://doi.org/10.1038/nrclinonc.2011.67.

(10)   Weilbaecher, K. N.; Guise, T. A.; McCauley, L. K. Cancer to Bone: A Fatal Attraction. *Nat. Rev. Cancer* **2011**, *11* (6), 411–425. https://doi.org/10.1038/nrc3055.

(11)   Mundy, G. R. Metastasis to Bone: Causes, Consequences and Therapeutic Opportunities. *Nat. Rev. Cancer* **2002**, *2* (8), 584–593. https://doi.org/10.1038/nrc867.

(12)   Maria, S. M.; Prukner, C.; Sheikh, Z.; Müller, F. A.; Komarova, S. V.; Barralet, J. E. Characterization of Biomimetic Calcium Phosphate Labeled with Fluorescent Dextran for Quantification of Osteoclastic Activity. *Acta Biomater.* **2015**, *20*, 140–146. https://doi.org/10.1016/j.actbio.2015.03.026.

(13)   Zhang, Z.; Egaña, J. T.; Reckhenrich, A. K.; Schenck, T. L.; Lohmeyer, J. A.; Schantz, J. T.; MacHens, H. G.; Schilling, A. F. Cell-Based Resorption Assays for Bone Graft Substitutes. *Acta Biomater.* **2012**, *8* (1), 13–19. https://doi.org/10.1016/j.actbio.2011.09.020.

(14)   Owen, R.; Reilly, G. C. In Vitro Models of Bone Remodelling and Associated Disorders. *Front. Bioeng. Biotechnol.* **2018**, *6* (OCT), 1–22. https://doi.org/10.3389/fbioe.2018.00134.

(15)  Kohli, N.; Ho, S.; Brown, S. J.; Sawadkar, P.; Sharma, V.; Snow, M.; García-gareta, E. Bone Remodelling in Vitro : Where Are We Headed ? -A Review on the Current Understanding of Physiological Bone Remodelling and in Fl Ammation and the Strategies for Testing Biomaterials in Vitro. *Bone* **2018**, *110*, 38–46. https://doi.org/10.1016/j.bone.2018.01.015.

(16)  Winkler, T.; Hoenig, E.; Gildenhaar, R.; Berger, G.; Fritsch, D.; Janssen, R.; Morlock, M. M.; Schilling, A. F. Volumetric Analysis of Osteoclastic Bioresorption of Calcium Phosphate Ceramics with Different Solubilities. *Acta Biomater.* **2010**, *6* (10), 4127–4135. https://doi.org/10.1016/j.actbio.2010.04.015.

(17)  Jeon, O. H.; Jeong, S. H.; Yoo, Y. M.; Kim, K. H.; Yoon, D. S.; Kim, C. H. Quantification of Temporal Changes in 3D Osteoclastic Resorption Pit Using Confocal Laser Scanning Microscopy. *Tissue Eng. Regen. Med.* **2012**, *9* (1), 29–35. https://doi.org/10.1007/s13770-012-0029-1.

(18)  Winkler, T.; Hoenig, E.; Huber, G.; Janssen, R.; Fritsch, D.; Gildenhaar, R.; Berger, G.; Morlock, M. M.; Schilling, A. F. Osteoclastic Bioresorption of Biomaterials: Two- and Three-Dimensional Imaging and Quantification. *Int. J. Artif. Organs* **2010**, *33* (4), 198–203. https://doi.org/10.1177/039139881003300404.

(19)  Rumpler, M.; Würger, T.; Roschger, P.; Zwettler, E.; Sturmlechner, I.; Altmann, P.; Fratzl, P.; Rogers, M. J.; Klaushofer, K. Osteoclasts on Bone and Dentin in Vitro: Mechanism of Trail Formation and Comparison of Resorption Behavior. *Calcif. Tissue Int.* **2013**, *93* (6), 526–539. https://doi.org/10.1007/s00223-013-9786-7.

(20)  Merrild, D. M. H.; Pirapaharan, D. C.; Andreasen, C. M.; Kjærsgaard-Andersen, P.; Møller, A. M. J.; Ding, M.; Delaissé, J. M.; Søe, K. Pit- and Trench-Forming Osteoclasts: A Distinction That Matters. *Bone Res.* **2015**, *3* (September), 1–11. https://doi.org/10.1038/boneres.2015.32.

(21) Patntirapong, S.; Habibovic, P.; Hauschka, P. V. Effects of Soluble Cobalt and Cobalt Incorporated into Calcium Phosphate Layers on Osteoclast Differentiation and Activation. *Biomaterials* **2009**, *30* (4), 548–555. https://doi.org/10.1016/j.biomaterials.2008.09.062.

(22) Yang, L.; Perez-Amodio, S.; Barrère-de Groot, F. Y. F.; Everts, V.; van Blitterswijk, C. A.; Habibovic, P. The Effects of Inorganic Additives to Calcium Phosphate on in Vitro Behavior of Osteoblasts and Osteoclasts. *Biomaterials* **2010**, *31* (11), 2976–2989. https://doi.org/10.1016/j.biomaterials.2010.01.002.

(23) Costa, D. O.; Prowse, P. D. H.; Chrones, T.; Sims, S. M.; Hamilton, D. W.; Rizkalla, A. S.; Dixon, S. J. The Differential Regulation of Osteoblast and Osteoclast Activity Bysurface Topography of Hydroxyapatite Coatings. *Biomaterials* **2013**, *34* (30), 7215–7226. https://doi.org/10.1016/j.biomaterials.2013.06.014.

(24) Deguchi, T.; Alanne, M. H.; Fazeli, E.; Fagerlund, K. M.; Pennanen, P.; Lehenkari, P.; Hänninen, P. E.; Peltonen, J.; Näreoja, T. In Vitro Model of Bone to Facilitate Measurement of Adhesion Forces and Super-Resolution Imaging of Osteoclasts. *Sci. Rep.* **2016**, *6* (March), 1–13. https://doi.org/10.1038/srep22585.

(25) van Gestel, N. A. P.; Schuiringa, G. H.; Hennissen, J. H. P. H.; Delsing, A. C. A.; Ito, K.; van Rietbergen, B.; Arts, J. J.; Hofmann, S. Resorption of the Calcium Phosphate Layer on S53P4 Bioactive Glass by Osteoclasts. *J. Mater. Sci. Mater. Med.* **2019**, *30* (8). https://doi.org/10.1007/s10856-019-6295-x.

(26) Siebers, M. C.; Matsuzaka, K.; Walboomers, X. F.; Leeuwenburgh, S. C. G.; Wolke, J. G. C.; Jansen, J. A. Osteoclastic Resorption of Calcium Phosphate Coatings Applied with Electrostatic Spray Deposition (ESD), in Vitro. *J. Biomed. Mater. Res. - Part A* **2005**, *74* (4), 570–580. https://doi.org/10.1002/jbm.a.30332.

(27)  Gross, K. A.; Muller, D.; Lucas, H.; Haynes, D. R. Osteoclast Resorption of Thermal Spray Hydoxyapatite Coatings Is Influenced by Surface Topography. *Acta Biomater.* **2012**, *8* (5), 1948–1956. https://doi.org/10.1016/j.actbio.2012.01.023.

(28)  Rao, H.; Tan, J.; Faruqi, F.; Beltzer, J.; Park, S. Corning ® Osteo Assay Surface : A New Tool to Study Osteoclast and Osteoblast Differentiation and Function. No. 3.

(29)  Bergsma, A.; Ganguly, S. S.; Wiegand, M. E.; Dick, D.; Williams, B. O.; Miranti, C. K. Regulation of Cytoskeleton and Adhesion Signaling in Osteoclasts by Tetraspanin CD82. *Bone Reports* **2019**, *10* (June 2018), 100196. https://doi.org/10.1016/j.bonr.2019.100196.

(30)  Crasto, G.; Kartner, N.; Manolson, M. F. Drug Screening Using Corning ® Osteo Assay Surface Customer Application Note. **2010**, No. 12571.

(31)  Park, S.; Osteoclast, H.; Cells, P. Primary Human Osteoclast Differentiation and Function on Dentine Discs and Corning ® Osteo Assay Surface Sn APPS Hots Materials and Methods :

(32)  Wood, R. M.; Rothenberg, M. Corning® Osteo Assay Surface 24 Well Plates with Transwell® Permeable Supports – A Useful Tool for Co-Culture Studies. **2012**, No. 2, 5–8.

(33)  Boeyens, J. C. A.; Deepak, V.; Chua, W.; Kruger, M. C.; Joubert, A. M.; Coetzee, M. Effects of Ω3- and Ω6-Polyunsaturated Fatty Acids on RANKL-Induced Osteoclast Differentiation of RAW264.7 Cells: A Comparative in Vitro Study. **2014**, 2584–2601. https://doi.org/10.3390/nu6072584.

(34)  Zeng, M.; Kim, Y. Y.; Anduix-Canto, C.; Frontera, C.; Laundy, D.; Kapur, N.; Christenson, H. K.; Meldrum, F. C. Confinement Generates Single-Crystal Aragonite Rods at Room Temperature. *Proc. Natl. Acad. Sci. U. S. A.* **2018**, *115* (30), 7670–7675. https://doi.org/10.1073/pnas.1718926115.

(35)  Kim, Y. Y.; Hetherington, N. B. J.; Noel, E. H.; Kröger, R.; Charnock, J. M.; Christenson, H. K.; Meldrum, F. C. Capillarity Creates Single-Crystal Calcite Nanowires from Amorphous Calcium Carbonate. *Angew. Chemie - Int. Ed.* **2011**, *50* (52), 12572–12577. https://doi.org/10.1002/anie.201104407.

(36) Wada, N.; Horiuchi, N.; Nishio, M.; Nakamura, M.; Nozaki, K.; Nagai, A.; Hashimoto, K.; Yamashita, K. Crystallization of Calcium Phosphate in Agar Hydrogels in the Presence of Polyacrylic Acid under Double Diffusion Conditions. *Cryst. Growth Des.* **2017**, *17* (2), 604–611. https://doi.org/10.1021/acs.cgd.6b01453.

(37) Cantaert, B.; Beniash, E.; Meldrum, F. C. The Role of Poly(Aspartic Acid) in the Precipitation of Calcium Phosphate in Confinement. *J. Mater. Chem. B* **2013**, *1* (48), 6586–6595. https://doi.org/10.1039/c3tb21296c.

(38) Loste, E.; Meldrum, F. C. Control of Calcium Carbonate Morphology by Transformation of an Amorphous Precursor in a Constrained Volume. *Chem. Commun.* **2001**, *1* (10), 901–902. https://doi.org/10.1039/b101563j.

(39) Schenk, A. S.; Albarracin, E. J.; Kim, Y. Y.; Ihli, J.; Meldrum, F. C. Confinement Stabilises Single Crystal Vaterite Rods. *Chem. Commun.* **2014**, *50* (36), 4729–4732. https://doi.org/10.1039/c4cc01093k.

(40) Cantaert, B.; Beniash, E.; Meldrum, F. C. Nanoscale Confinement Controls the Crystallization of Calcium Phosphate: Relevance to Bone Formation. *Chem. - A Eur. J.* **2013**, *19* (44), 14918–14924. https://doi.org/10.1002/chem.201302835.

(41) Meldrum, F. C.; O'Shaughnessy, C. Crystallization in Confinement. *Adv. Mater.* **2020**, *32* (31). https://doi.org/10.1002/adma.202001068.

(42) Duan, K.; Tang, A.; Wang, R. A New Evaporation-Based Method for the Preparation of Biomimetic Calcium Phosphate Coatings on Metals. *Mater. Sci. Eng. C* **2009**, *29* (4), 1334–1337. https://doi.org/10.1016/j.msec.2008.10.028.

(43) Garbuz, D. S.; Hu, Y.; Kim, W. Y.; Duan, K.; Masri, B. A.; Oxland, T. R.; Burt, H.; Wang, R.; Duncan, C. P. Enhanced Gap Filling and Osteoconduction Associated with Alendronate-Calcium Phosphate-Coated Porous Tantalum. *J. Bone Jt. Surg. - Ser. A* **2008**, *90* (5), 1090–1100. https://doi.org/10.2106/JBJS.G.00415.

(44)  DUAN, K.; HU, Y.; LONG, K.; TOMS, A.; BURT, H. M.; OXLAND, T. R.; MASRI, B. A.; DUNCAN, C. P.; GARBUZ, D. S.; WANG, R. Effect of Alendronate-Containing Coatings on Osteointegration Into Porous Tantalum in a Cortical Bone Model. *Nano Life* **2012**, *02* (01), 1250007. https://doi.org/10.1142/S1793984411000414.

(45)  Lu, Y.; Cai, Z.; Xiao, G.; Keller, E. T.; Mizokami, A.; Yao, Z.; Roodman, G. D.; Zhang, J. Monocyte Chemotactic Protein-1 Mediates Prostate Cancer-Induced Bone Resorption. *Cancer Res.* **2007**, *67* (8), 3646–3653. https://doi.org/10.1158/0008-5472.CAN-06-1210.

(46)  Alsulaiman, M.; Bais, M. V.; Trackman, P. C. Lysyl Oxidase Propeptide Stimulates Osteoblast and Osteoclast Differentiation and Enhances PC3 and DU145 Prostate Cancer Cell Effects on Bone in Vivo. *J. Cell Commun. Signal.* **2016**, *10* (1), 17–31. https://doi.org/10.1007/s12079-015-0311-9.

(47)  Hayman, A. Tartrate-Resistant Acid Phosphatase (TRAP) and the Osteoclast/Immune Cell Dichotomy. *Autoimmunity* **2008**, *41* (3), 218–223. https://doi.org/10.1080/08916930701694667.

(48)  Zhang, Y.; Chen, S. E.; Shao, J.; Van Den Beucken, J. J. J. P. Combinatorial Surface Roughness Effects on Osteoclastogenesis and Osteogenesis. *ACS Appl. Mater. Interfaces* **2018**, *10* (43), 36652–36663. https://doi.org/10.1021/acsami.8b10992.

(49)  Arellano-Jiménez, M. J.; García-García, R.; Reyes-Gasga, J. Synthesis and Hydrolysis of Octacalcium Phosphate and Its Characterization by Electron Microscopy and X-Ray Diffraction. *J. Phys. Chem. Solids* **2009**, *70* (2), 390–395. https://doi.org/10.1016/j.jpcs.2008.11.001.

(50)  Suzuki, O. Octacalcium Phosphate: Osteoconductivity and Crystal Chemistry. *Acta Biomater.* **2010**, *6* (9), 3379–3387. https://doi.org/10.1016/j.actbio.2010.04.002.

(51)  An, I. R. E. H. M.; Mary, Q.; Road, M. E. Characterization of Hydroxyapatite and Carbonated Apatite by Photo Acoustic FTIR Spectroscopy. *8*, 1–4.

(52)  Koutsopoulos, S. Synthesis and Characterization of Hydroxyapatite Crystals : A Review Study on the Analytical Methods. **2002**, No. February, 31–34.

(53)  Mater, C.; Markovic, M.; Fowler, B. Octacalcium Phosphate Carboxylates . 2 . Characterization and Structural Considerations. **1993**, *18* (7), 1406–1416. https://doi.org/10.1021/cm00034a008.

(54)  Fowler, B. O.; Markovic, M.; Brown, W. E. Octacalcium Phosphate .3. Infrared and Raman Vibrational-Spectra. *Chem. Mater.* **1993**, *5* (c), 1417–1423. https://doi.org/10.1021/cm00034a009.

(55)  Vincent, C.; Kogawa, M.; Findlay, D. M.; Atkins, G. J. The Generation of Osteoclasts from RAW 264.7 Precursors in Defined, Serum-Free Conditions. *J. Bone Miner. Metab.* **2009**, *27* (1), 114–119. https://doi.org/10.1007/s00774-008-0018-6.

(56)  Ng, A. Y.; Tu, C.; Shen, S.; Xu, D.; Oursler, M. J.; Qu, J.; Yang, S. Comparative Characterization of Osteoclasts Derived From Murine Bone Marrow Macrophages and RAW 264.7 Cells Using Quantitative Proteomics. *JBMR Plus* **2018**, *2* (6), 328–340. https://doi.org/10.1002/jbm4.10058.

(57)  Kong, L.; Smith, W.; Hao, D. Overview of RAW264.7 for Osteoclastogensis Study: Phenotype and Stimuli. *J. Cell. Mol. Med.* **2019**, *23* (5), 3077–3087. https://doi.org/10.1111/jcmm.14277.

(58)  Loste, E.; Park, R. J.; Warren, J.; Meldrum, F. C. Precipitation of Calcium Carbonate in Confinement. *Adv. Funct. Mater.* **2004**, *14* (12), 1211–1220. https://doi.org/10.1002/adfm.200400268.

(59)  Ihli, J.; Wang, Y.; Cantaert, B.; Kim, Y.; Green, D. C.; Bomans, P. H. H.; Sommerdijk, N. A. J. M.; Meldrum, F. C. Precipitation of Amorphous Calcium Oxalate in Aqueous Solution. **2015**. https://doi.org/10.1021/acs.chemmater.5b01642.

(60)  Wang, L.; Nancollas, G. H. Calcium Orthophosphates: Crystallization and Dissolution. *Chem. Rev.* **2008**, *108* (11), 4628–4669. https://doi.org/10.1021/cr0782574.

(61)  Ihli, J.; Bots, P.; Kulak, A.; Benning, L. G.; Meldrum, F. C. Elucidating Mechanisms of Diffusion-Based Calcium Carbonate Synthesis Leads to Controlled Mesocrystal Formation. *Adv. Funct. Mater.* **2013**, *23* (15), 1965–1973. https://doi.org/10.1002/adfm.201201742.

(62)  Tovani, C. B.; Oliveira, T. M.; Soares, M. P. R.; Nassif, N.; Fukada, S. Y.; Ciancaglini, P.; Gloter, A.; Ramos, A. P. Strontium Calcium Phosphate Nanotubes as Bioinspired Building Blocks for Bone

Regeneration. *ACS Appl. Mater. Interfaces* **2020**, *12* (39), 43422–43434. https://doi.org/10.1021/acsami.0c12434.

(63) H, Q.; Z, J.; T, H.; B, U. Growth Process and Crystallographic Properties of Ammonia-Induced Vaterite. **2012**, *97*, 1437–1445.

(64) Rogers, M. J.; Mönkkönen, J.; Munoz, M. A. Molecular Mechanisms of Action of Bisphosphonates and New Insights into Their Effects Outside the Skeleton. *Bone* **2020**, *139* (April), 115493. https://doi.org/10.1016/j.bone.2020.115493.

(65) Faucheux, C.; Verron, E.; Soueidan, A.; Josse, S.; Arshad, M. D.; Janvier, P.; Pilet, P.; Bouler, J. M.; Bujoli, B.; Guicheux, J. Controlled Release of Bisphosphonate from a Calcium Phosphate Biomaterial Inhibits Osteoclastic Resorption in Vitro. *J. Biomed. Mater. Res. - Part A* **2009**, *89* (1), 46–56. https://doi.org/10.1002/jbm.a.31989.

(66) Ng, P. Y.; Patricia Ribet, A. B.; Pavlos, N. J. Membrane Trafficking in Osteoclasts and Implications for Osteoporosis. *Biochem. Soc. Trans.* **2019**, *47* (2), 639–650. https://doi.org/10.1042/BST20180445.

(67) Kaighn, M. E.; Narayan, K. S.; Ohnuki, Y.; Lechner, J. F.; Jones, L. W. Establishment and Characterization of a Human Prostatic Carcinoma Cell Line (PC-3). *Invest. Urol.* **1979**.

(68) Araujo, J. C.; Poblenz, A.; Corn, P.; Parikh, N. U.; Starbuck, M. W.; Thompson, J. T.; Lee, F.; Logothetis, C. J.; Darnay, B. G. Dasatinib Inhibits Both Osteoclast Activation and Prostate Cancer PC-3 Cell-Induced Osteoclast Formation. *Cancer Biol. Ther.* **2009**, *8* (22), 2153–2159. https://doi.org/10.4161/cbt.8.22.9770.

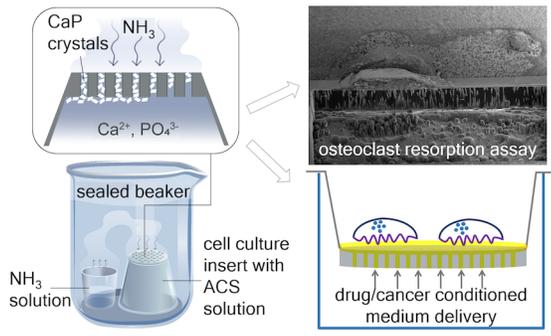

CaP crystals
NH₃
Ca²⁺, PO₄³⁻
sealed beaker
cell culture insert with ACS solution
NH₃ solution

osteoclast resorption assay

drug/cancer conditioned medium delivery

For Table of Contents Only